%% file: Paper_Submit.tex
\documentclass[letterpaper,english,reprint, aps]{revtex4-1}
\usepackage[T1]{fontenc}
\usepackage[latin9]{inputenc}
\usepackage{amsmath}
\usepackage{amssymb}

\makeatletter

\pdfpageheight\paperheight
\pdfpagewidth\paperwidth

\makeatother

\usepackage{babel}
\begin{document}
\title{Formulation of the Generator Coordinate Method with arbitrary bases}
\author{L.M. Robledo}
\email{luis.robledo@uam.es}

\affiliation{Departamento de Física Teórica and CIAF, Universidad Autónoma de Madrid,
E-28049 Madrid, Spain}
\affiliation{Center for Computational Simulation, Universidad Polit\textbackslash 'ecnica
de Madrid, Campus de Montegancedo, Boadilla del Monte, E-28660-Madrid,
Spain}
\date{\today}
\begin{abstract}
The existing formalism used to compute the operator overlaps necessary
to carry out generator coordinate method calculations using a set
of Hartree- Fock- Bogoliubov wave functions, is generalized to the
case where each of the HFB states are expanded in different arbitrary
bases spanning different sub-space of the Hilbert space. 
\end{abstract}
\maketitle

\section{\label{sec:Intro}Introduction}

The calculation of operator overlap between general Hartree- Fock
(HF or Slater) or Hartree- Fock- Bogoliubov (HFB) mean field wave
functions is a common task in many physics areas like nuclear physics
\citep{Sheikh2021}, condensed matter \citep{Sheikh2021} or quantum
chemistry \citep{Piela2020}. It is required in the restoration of
spontaneously broken ( by the mean field) symmetries or in the consideration
of fluctuations beyond the mean field in the context of the configuration
interaction (CI) or the generator coordinate method (GCM) \citep{ring2000,bender2003,Robledo2019,Sheikh2021}.
In both cases, linear combinations of mean field wave functions of
the HF or HFB type are used to build a variational space. The set
of HFB wave functions is usually chosen as to explore the corner of
the Hilbert space relevant to the physics to be described or it is
dictated by the symmetry to be restored. The evaluation of the overlaps
is greatly simplified by using the generalized Wick theorem (GWT)
for general HFB states \citep{Onishi1966367,Balian.69} or its equivalent
for Slater determinants \citep{PhysRev.97.1490}. Generalizations
to consider different peculiarities in the calculations of the overlaps
have been developed along the years both at zero \citep{PhysRevC.50.2874,PhysRevC.79.021302,PhysRevC.84.014307,PhysRevLett.108.042505,RodLag2020}
or finite temperature \citep{Gau60,Per07,(Ros94)}. The GWT implicitly
assumes that all the quasiparticle operators of the Bogoliubov transformation
are expanded in a common basis that is taken often as finite dimensional
due to computational complexity reasons. However, in many practical
applications the bases to be used for each of the HFB states have
a different set of parameters (for instance, oscillator lengths in
the harmonic oscillator basis case) or, in the context of symmetry
restoration, the basis is not closed under the symmetry operation
(for instance, an arbitrary translation of the HO basis). The most
straightforward solution to this problem is to use a common basis
(with the same oscillator lengths) for all the states of the HFB set
or, in the case of symmetry restoration, a basis which is closed under
the symmetry operation (HO basis with the same oscillator lengths
along the three spatial directions in the case of rotations, a plane
wave basis in the case of translations, etc). However, if the use
of a localized basis is required along with spatial translations,
the only easy strategy is to use very big basis and to carefully check
the convergence of the results with basis size \citep{Schmid2001,Rod04a}.
These simple strategies come to a cost, namely, to increase the basis
size and therefore the computational complexity. The situation is
specially delicate, for instance, in fission studies where the very
broad range of nuclear shapes to be considered in the fission process
makes impractical to use a basis with equal oscillator lengths (in
fact, all practitioners of fission using either one center or two
center HO basis often use different, optimized basis parameters for
each quadrupole moment defining the fission process) \citep{Schunck2016,Marevic2020}.
At this point the reader might wonder why not to do the calculation
in the mesh. This solution is however impractical in general and it
is only useful for zero range interactions with trivial local exchange
terms. In addition, the action of the symmetry operators in the mesh
requires of assumptions and approximations in the realization of the
generators of the symmetry that have to be carefully considered \citep{Baye1984}.
Therefore, the only viable solution to all the problems with non-complete
bases relies on the formal extension of the original basis as to make
it complete with the added states having zero occupancy. This approach
has been pursued in Refs \citep{BONCHE1990466,Valor2000145} for unitary
and in Ref \citep{PhysRevC.50.2874} for general canonical transformations.
However, in those references it is not clear whether one can compute
the overlaps in terms of quantities defined in the starting, finite
size, bases. The purpose of this paper is to extend the formalism
of \citep{PhysRevC.50.2874} to prove that the overlaps can always
be obtained in terms of what we will call intrinsic quantities (i.e.
quantities that are defined solely in the given finite bases) and
therefore there is no need to refer to the complementary (often infinite-dimensional)
sub-space required to make the bases complete. In addition, by using
the Lower-Upper (LU) decomposition of the overlap matrix, it will
be possible to express all the different quantities in a more familiar
form facilitating the application of the obtained formulas. The application
of the formalism to the use of harmonic oscillator wave functions
with different oscillator lengths or the more general case involving
rotated and translated basis is deferred to future publications.

\section{The generalized Wick theorem for arbitrary basis}

The goal is to evaluate the overlap of general multi-body operators
between arbitrary HFB wave functions
\begin{equation}
\frac{\langle\phi_{0}|\hat{O}|\phi_{1}\rangle}{\langle\phi_{0}|\phi_{1}\rangle}\label{eq:over1}
\end{equation}
where each of the HFB states are expanded in different bases not connected
by unitary transformations (i.e. not expanding the same subspace of
the whole Hilbert space). We will denote the corresponding bases and
associated creation operators as $\mathcal{B}_{0}=\{c_{0,k}^{\dagger},k=1,\ldots,N_{0}\}$
in the case of $|\phi_{0}\rangle$ and $\mathcal{B}_{1}=\{c_{1,k}^{\dagger},k=1,\ldots,N_{1}\}$
in the case of $|\phi_{1}\rangle$. It is implicitly assumed that
fermion canonical anticommutation relations (CAR) are preserved among
each basis set, i.e $\{c_{i,k},c_{i,k'}\}=\delta_{kk'}$ but there
is an overlap matrix connecting both sets $\{c_{0,k}^{\dagger},c_{1,l}\}=_{0}\langle k|l\rangle_{1}=\mathcal{R}_{kl}$.
For simplicity, we will consider in the following $N_{0}=N_{1}=N$,
but note that the most general case can be easily accommodated in
the formalism. We will also introduce the complement of the two bases
$\bar{\mathcal{B}}_{0}=\{c_{0,k}^{\dagger},k=N+1,\ldots,\infty\}$
and $\bar{\mathcal{B}}_{1}=\{c_{1,k}^{\dagger},k=N+1,\ldots,\infty\}$
such that $\mathcal{B}_{0}\cup\bar{\mathcal{B}}_{0}=\{c_{0,k}^{\dagger}\}^{\infty}$
and $\mathcal{B}_{1}\cup\bar{\mathcal{B}}_{1}=\{c_{1,k}^{\dagger}\}^{\infty}$
expand the whole separable Hilbert space and therefore represent bases
connected by a unitary transformation matrix $R$ (not to be confused
with $\mathcal{R}$). We are assuming separable Hilbert spaces for
which a countable orthonormal bases exist and therefore the introduction
of a (infinite dimensional) matrix $R$ makes sense. Let us also introduce
the quasiparticle annihilation operators $\alpha_{i\mu}$ ($i=0,1)$,
which annihilate $|\phi_{i}\rangle$, and are written in terms of
the complete bases $\{c_{i,k}^{\dagger}\}^{\infty}$ through the standard
definition
\[
\alpha_{i\mu}=\sum_{k}\left(U_{i}^{*}\right)_{k\mu}c_{i,k}+\left(V_{i}^{*}\right)_{k\mu}c_{i,k}^{\dagger}.
\]
By using the following block structure for the Bogoliubov amplitudes
$U_{i}$ and $V_{i}$ 
\begin{equation}
V_{i}=\left(\begin{array}{cc}
\bar{V}_{i} & 0\\
0 & 0
\end{array}\right),\;\;U_{i}=\left(\begin{array}{cc}
\bar{U}_{i} & 0\\
0 & d_{i}
\end{array}\right),\label{eq:block}
\end{equation}
where $\bar{V}_{i}$and $\bar{U}_{i}$ are $N\times N$ matrices,
we can accommodate into the formalism the set of $N$ quasiparticle
operators $\alpha_{i\mu}$ with $\mu=1,\ldots,N$, corresponding to
the quasiparticle operators expanded in the truncated bases $\mathcal{B}_{i}$.
The $d_{i}$ are arbitrary unitary matrices that should not appear
explicitly in the final expressions. It is also convenient to express
the unitary matrix $R$ connecting $\mathcal{B}_{0}\cup\bar{\mathcal{B}}_{0}$
and $\mathcal{B}_{1}\cup\bar{\mathcal{B}}_{1}$ as a block matrix
\[
R=\left(\begin{array}{cc}
\mathcal{R} & \mathcal{S}\\
\mathcal{T} & \mathcal{U}
\end{array}\right).
\]
The matrix $R$ is just the representation of the unitary operator
$\hat{\mathcal{T}}_{01}$ connecting the two complete  bases 
\[
\hat{\mathcal{T}}_{01}c_{0,k}^{\dagger}\hat{\mathcal{T}}_{01}^{\dagger}=c_{1,k}^{\dagger}.
\]
The $\hat{\mathcal{T}}_{01}$ operator can be a symmetry operator
like a spatial translation, a rotation or the dilatation operator
when dealing with HO bases differing in their oscillator lengths.
In all the cases (and this is an implicit requirement of the present
development) the operator is the exponential of an one-body operator.
Finally, let us introduce the HFB state $|\tilde{\phi}_{1}\rangle$
and the associated annihilation operators $\tilde{\alpha}_{1,\mu}$
defined by the relations
\[
\hat{\mathcal{T}}_{01}|\tilde{\phi}_{1}\rangle=|\phi_{1}\rangle
\]
and
\[
\hat{\mathcal{T}}_{01}\tilde{\alpha}_{1,\mu}\hat{\mathcal{T}}_{01}^{\dagger}=\alpha_{1,\mu}.
\]
The annihilation operators $\tilde{\alpha}_{1,\mu}$ share the Bogoliubov
amplitudes with $\alpha_{1,\mu}$ but are expressed in the basis $\mathcal{B}_{0}$
\[
\tilde{\alpha}_{1\mu}=\sum_{k=1}^{N}\left(\bar{U}_{1}^{*}\right)_{k\mu}c_{0,k}+\left(\bar{V}_{1}^{*}\right)_{k\mu}c_{0,k}^{\dagger}.
\]
Let us also introduce the $\hat{\mathcal{T}}_{B}$ operator of the
Bogoliubov transformation from $\alpha_{0,\mu}$ to $\tilde{\alpha}_{1,\mu}$
\[
\hat{\mathcal{T}}_{B}\alpha_{0,\mu}\hat{\mathcal{T}}_{B}^{+}=\tilde{\alpha}_{1,\mu}
\]
and 
\[
\hat{\mathcal{T}}_{B}|\phi_{0}\rangle=|\tilde{\phi}_{1}\rangle
\]
To compute the overlap of Eq. (\ref{eq:over1}) it will prove convenient
to write the operator $\hat{O}$ in terms of both bases $\{c_{0,k}^{\dagger}\}^{\infty}$
and $\{c_{1,k}^{\dagger}\}^{\infty}$ in a convenient way. For instance,
for a two-body operator we will use
\begin{equation}
\hat{\upsilon}=\frac{1}{4}\sum_{k_{1}k_{2}l_{1}l_{2}}\tilde{\upsilon}_{k_{1}k_{2}l_{1}l_{2}}^{01}c_{0k_{1}}^{\dagger}c_{0,k_{2}}^{\dagger}c_{1,l_{2}}c_{1,l_{1}}\label{eq:Twobody}
\end{equation}
where the antisymmetrized two-body matrix element is given by $\tilde{\upsilon}_{k_{1}k_{2}l_{1}l_{2}}^{01}=\upsilon_{k_{1}k_{2}l_{1}l_{2}}^{01}-\upsilon_{k_{1}k_{2}l_{2}l_{1}}^{01}$with
\begin{equation}
\upsilon_{k_{1}k_{2}l_{2}l_{1}}^{01}=_{0}\langle k_{1}k_{2}|\hat{\upsilon}|l_{1}l_{2}\rangle_{1}\label{eq:v01me}
\end{equation}
the interaction's overlap matrix elements. The sums in Eq (\ref{eq:Twobody})
extend over the complete bases $\{c_{0,k}^{\dagger}\}^{\infty}$ or
$\{c_{1,k}^{\dagger}\}^{\infty}$ to faithfully represent the operators.
The advantage of Eq (\ref{eq:Twobody}) is that the annihilation operators
acting on $|\phi_{1}\rangle$ lead to a linear combination of multi-quasiparticle
excitations which are all of them expressed in terms of basis $\mathcal{B}_{1}$
alone, whereas the creation operators action to the left on $|\phi_{0}\rangle$
will do the same but in terms of $\mathcal{B}_{0}$. This is the key
point to obtain expression for the overlaps depending solely in the
bases used (and not their complements). The overlaps are computed
by transforming to the quasiparticle representation and applying GWT.
With the previous considerations we have to evaluate
\begin{align}
\frac{\langle\phi_{0}|\alpha_{0,\mu_{1}}\ldots\alpha_{0,\mu_{M}}\alpha_{1,\nu_{M}}^{+}\ldots\alpha_{1,\nu_{1}}^{+}|\phi_{1}\rangle}{\langle\phi_{0}|\phi_{1}\rangle} & =\\
\frac{\langle\phi_{0}|\alpha_{0,\mu_{1}}\ldots\alpha_{0,\mu_{M}}\hat{\mathcal{T}}\alpha_{0,\nu_{M}}^{+}\ldots\alpha_{0,\nu_{1}}^{+}|\phi_{0}\rangle}{\langle\phi_{0}|\hat{\mathcal{T}}|\phi_{0}\rangle}\label{eq:twobme}
\end{align}
with $\hat{\mathcal{T}}=\hat{\mathcal{T}}_{01}\hat{\mathcal{T}}_{B}$
the product of exponential of one-body operators that can also be
written as the exponential of an one-body operator \citep{Balian.69}.
To evaluate these overlaps we will make heavy use of the results of
Ref \citep{PhysRevC.50.2874} (denoted I hereafter). The main difference
between the present results and those in I is that there we considered
$\langle\phi_{0}|\hat{A}\hat{\mathcal{T}}|\phi_{0}\rangle/\langle\phi_{0}|\hat{\mathcal{T}}|\phi_{0}\rangle,$
instead of having $\hat{\mathcal{T}}$ ``in the middle'' of $\hat{A}$.
Fortunately, we can use the decomposition given in Eq (I.39) $\hat{\mathcal{T}}=\hat{\mathcal{T}}_{1}\hat{\mathcal{T}}_{2}\hat{\mathcal{T}}_{3}\left(\det R\right)^{1/2}$
(see also \citep{Balian.69}) where each of the $\hat{\mathcal{T}_{i}}$
can be decomposed in turn as the product of three elementary transformations
$\hat{\mathcal{T}_{i}}=\hat{\mathcal{T}_{i}}^{20}\hat{\mathcal{T}_{i}}^{11}\hat{\mathcal{T}_{i}}^{02}\mathcal{T}_{i}^{0}$
where the $\mathcal{T}_{i}^{nm}$ represents the exponential of an
one-body operator expressed as linear combinations of the product
of $n$ quasiparticle creation ($\alpha_{0,\mu}^{+}$) and m annihilation
operators ($\alpha_{0,\mu}$) and $\mathcal{T}_{i}^{0}$ represents
a constant factor. According to Eqs (42-54) in I we have $\hat{\mathcal{T}_{1}}^{02}=\hat{\mathcal{T}_{3}}^{20}=\mathbb{I}$
and $\mathcal{T}_{1}^{0}=\mathcal{T}_{3}^{0}=1$ which allows to define
the operators
\begin{equation}
\hat{\mathcal{T}}_{L}=\hat{\mathcal{T}_{1}}^{20}\hat{\mathcal{T}_{1}}^{11}\hat{\mathcal{T}_{2}}^{20}\hat{\mathcal{T}_{2}}^{11}\label{eq:TL}
\end{equation}
and
\begin{equation}
\hat{\mathcal{T}}_{R}=\hat{\mathcal{T}_{2}}^{02}\hat{\mathcal{T}_{3}}^{11}\hat{\mathcal{T}_{3}}^{02}\label{eq:TR}
\end{equation}
such that $\hat{\mathcal{T}}=\hat{\mathcal{T}}_{L}\hat{\mathcal{T}}_{R}$
(up to an irrelevant $\mathcal{T}_{2}^{0}$ factor) and with the properties
$\langle\phi_{0}|\hat{\mathcal{T}}_{L}=\langle\phi_{0}|$ and $\hat{\mathcal{T}}_{R}|\phi_{0}\rangle=|\phi_{0}\rangle$.
We use now the operators $\hat{\mathcal{T}}_{L}$ and $\hat{\mathcal{T}}_{R}$
to define the quasiparticle operators (satisfying canonical anti-commutation
relations CARs) $d_{0}$, $\bar{d}_{0}$, $b_{0}$ and $\bar{b}_{0}$
by means of the following relations
\begin{equation}
\left(\begin{array}{c}
d_{0}\\
\bar{d}_{0}
\end{array}\right)=\hat{\mathcal{T}}_{L}^{-1}\left(\begin{array}{c}
\alpha_{0}\\
\alpha_{0}^{+}
\end{array}\right)\hat{\mathcal{T}}_{L}\label{eq:d0}
\end{equation}
\begin{equation}
\left(\begin{array}{c}
b_{0}\\
\bar{b}_{0}
\end{array}\right)=\hat{\mathcal{T}}_{R}\left(\begin{array}{c}
\alpha_{0}\\
\alpha_{0}^{+}
\end{array}\right)\hat{\mathcal{T}}_{R}^{-1}\label{eq:b0}
\end{equation}
we can finally express the matrix element of Eq (\ref{eq:twobme})
as the mean value 
\begin{equation}
\langle\phi_{0}|d_{0,\mu_{1}}\ldots d_{0,\mu_{M}}\bar{b}_{0,\nu_{M}}\ldots\bar{b}_{0,\nu_{1}}|\phi_{0}\rangle.\label{eq:FinalMV}
\end{equation}
The $d_{0}$ and $b_{0}$ are quasiparticle operators linear combinations
of the $\alpha_{0}$ and $\alpha_{0}^{+}$. Therefore, one can use
the standard Wick's theorem to evaluate Eq (\ref{eq:FinalMV}) in
terms of the contractions $\langle\phi_{0}|d_{0,\mu}\bar{b}_{0,\nu}|\phi_{0}\rangle$,
$\langle\phi_{0}|d_{0,\mu}d_{0,\nu}|\phi_{0}\rangle$ and $\langle\phi_{0}|\bar{b}_{0,\mu}\bar{b}_{0,\nu}|\phi_{0}\rangle$.
In order to obtain the expressions of the contractions we need the
explicit form of the $d_{0}$ and $\bar{b}_{0}$ operators in terms
of $\alpha_{0}$ and $\alpha_{0}^{+}$. Using Eqs (32), (A5) and (A6)
of I we get (using the notation of I)
\[
\left(\begin{array}{c}
d_{0}\\
\bar{d}_{0}
\end{array}\right)=\left(\begin{array}{cc}
D_{11} & D_{12}\\
0 & D_{22}
\end{array}\right)\left(\begin{array}{c}
\alpha_{0}\\
\alpha_{0}^{+}
\end{array}\right)
\]
with
\begin{align*}
\left(\begin{array}{cc}
D_{11} & D_{12}\\
0 & D_{22}
\end{array}\right) & =\left(\begin{array}{cc}
T_{11}^{(1)} & T_{12}^{(1)}\\
0 & T_{22}^{(1)}
\end{array}\right)\left(\begin{array}{cc}
\mathbb{I} & K^{(1)}\\
0 & \mathbb{I}
\end{array}\right)\\
 & \times\left(\begin{array}{cc}
e^{L^{(2)}} & 0\\
0 & e^{-\left(L^{(2)}\right)^{T}}
\end{array}\right).
\end{align*}
In the same way and using Eqs (35), (A7) and (51-54) of I we obtain
\[
\left(\begin{array}{c}
b_{0}\\
\bar{b}_{0}
\end{array}\right)=\left(\begin{array}{cc}
B_{11} & 0\\
B_{21} & B_{22}
\end{array}\right)\left(\begin{array}{c}
\alpha_{0}\\
\alpha_{0}^{+}
\end{array}\right).
\]
with
\begin{align*}
\left(\begin{array}{cc}
B_{11} & 0\\
B_{21} & B_{22}
\end{array}\right) & =\left(\begin{array}{cc}
\mathbb{I} & 0\\
-M^{(3)} & \mathbb{I}
\end{array}\right)\left(\begin{array}{cc}
e^{-L^{(3)}} & 0\\
0 & \left(e^{L^{(3)}}\right)^{T}
\end{array}\right)\\
 & \times\left(\begin{array}{cc}
\mathbb{I} & 0\\
-M^{(2)} & \mathbb{I}
\end{array}\right)
\end{align*}
The relevant contractions are easily obtained 
\begin{align}
\langle\phi_{0}|d_{0,\mu}\bar{b}_{0,\nu}|\phi_{0}\rangle & \equiv C_{\mu\nu}=\left(D_{11}B_{22}^{T}\right)_{\mu\nu}\label{eq:C}\\
\langle\phi_{0}|d_{0,\mu}d_{0,\nu}|\phi_{0}\rangle & \equiv D_{\mu\nu}=\left(D_{11}D_{12}^{T}\right)_{\mu\nu}\label{eq:D}\\
\langle\phi_{0}|\bar{b}_{0,\mu}\bar{b}_{0,\nu}|\phi_{0}\rangle & \equiv E_{\mu\nu}=\left(B_{21}B_{22}^{T}\right)_{\mu\nu}\label{eq:E}
\end{align}
Using the explicit form of the matrices $T_{ij}^{(1)}$, $M^{(2)}$,
$M^{(3)}$, $L^{(2)}$ and $L^{(3)}$ given in I and their block decomposition
in terms of the original basis and its complement one obtain the desired
expressions for the contractions. Using Eqs (32a), (47) and (52) of
I we arrive to
\[
\frac{\langle\phi_{0}|\alpha_{0,\mu}\alpha_{1,\nu}^{+}|\phi_{1}\rangle}{\langle\phi_{0}|\phi_{1}\rangle}=\langle\phi_{0}|d_{0,\mu}\bar{b}_{0,\nu}|\phi_{0}\rangle=\left(\begin{array}{cc}
\left(A^{T}\right)^{-1} & \bullet\\
\bullet & \mathbb{\bullet}
\end{array}\right)_{\mu\nu}
\]
Using Eqs (32a), (46) and (32b) of I we obtain
\[
\frac{\langle\phi_{0}|\alpha_{0,\mu}\alpha_{0,\nu}|\phi_{1}\rangle}{\langle\phi_{0}|\phi_{1}\rangle}\langle\phi_{0}|d_{0,\mu}d_{0,\nu}|\phi_{0}\rangle=\left(\begin{array}{cc}
-BA^{-1} & \bullet\\
\bullet & \mathbb{\bullet}
\end{array}\right)_{\mu\nu}
\]
Finally, using Eqs (48), (52) and (53) of I we get
\[
\frac{\langle\phi_{0}|\alpha_{1,\mu}^{+}\alpha_{1,\nu}^{+}|\phi_{1}\rangle}{\langle\phi_{0}|\phi_{1}\rangle}=\langle\phi_{0}|\bar{b}_{0,\mu}\bar{b}_{0,\nu}|\phi_{0}\rangle=\left(\begin{array}{cc}
-A^{-1}\bar{B} & \bullet\\
\bullet & \mathbb{\bullet}
\end{array}\right)_{\mu\nu}
\]
where the indices of the matrices $A$, $B$ and $\bar{B}$ (to be
defined below) run over the original space spanned by the original
bases and the symbol ``$\bullet$'' represents irrelevant matrices
defined in the complementary sub-spaces. The matrices $A$, $B$ and
$\bar{B}$ are defined through the relation
\begin{equation}
\left(\begin{array}{cc}
\bar{A} & B\\
\bar{B} & A
\end{array}\right)=\left(\begin{array}{cc}
\bar{U}_{0}^{\dagger} & \bar{V}_{0}^{\dagger}\\
\bar{V}_{0}^{T} & \bar{U}_{0}^{T}
\end{array}\right)\left(\begin{array}{cc}
\mathcal{R} & 0\\
0 & \left(\mathcal{R}^{T}\right)^{-1}
\end{array}\right)\left(\begin{array}{cc}
\bar{U}_{1} & \bar{V}_{1}^{*}\\
\bar{V}_{1} & \bar{U}_{1}^{*}
\end{array}\right).\label{eq:AB}
\end{equation}
With the above results is evident that the evaluation of the overlap
$\langle\phi_{0}|c_{0k_{1}}^{\dagger}c_{0,k_{2}}^{\dagger}c_{1,l_{2}}c_{1,l_{1}}|\phi_{1}\rangle/\langle\phi_{0}|\phi_{1}\rangle$
can be carried out according to the rules of the GWT in terms of the
contractions 
\begin{align}
\rho_{lk}^{01} & =\frac{\langle\phi_{0}|c_{0,k}^{\dagger}c_{1,l}|\phi_{1}\rangle}{\langle\phi_{0}|\phi_{1}\rangle}=\left(V_{1}^{*}C^{T}V_{0}^{T}\right)_{lk}\label{eq:rho01}\\
 & =\left(\begin{array}{cc}
\bar{V}_{1}^{*}\left(A\right)^{-1}\bar{V}_{0}^{T} & 0\\
0 & 0
\end{array}\right)\\
\bar{\kappa}_{k_{1}k_{2}}^{01} & =\frac{\langle\phi_{0}|c_{0,k_{1}}^{\dagger}c_{0,k_{2}}^{\dagger}|\phi_{1}\rangle}{\langle\phi_{0}|\phi_{1}\rangle}=\left(V_{0}U_{0}^{+}+V_{0}DV_{0}^{T}\right)_{k_{1}k_{2}}\label{eq:kappa01}\\
 & =\left(\begin{array}{cc}
\bar{V}_{0}\bar{U}_{0}^{+}-\bar{V}_{0}BA^{-1}\bar{V}_{0}^{T} & 0\\
0 & 0
\end{array}\right)\\
\kappa_{l_{1}l_{2}}^{10} & =\frac{\langle\phi_{0}|c_{1,l_{1}}c_{1,l_{2}}|\phi_{1}\rangle}{\langle\phi_{0}|\phi_{1}\rangle}=\left(U_{1}V_{1}^{+}+V_{1}^{*}EV_{1}^{+}\right)_{l_{1}l_{2}}\label{eq:kappa10}\\
 & =\left(\begin{array}{cc}
\bar{U}_{1}\bar{V}_{1}^{+}-\bar{V}_{1}^{*}A^{-1}\bar{B}\bar{V}_{1}^{+} & 0\\
0 & 0
\end{array}\right)
\end{align}
The result shows that the contractions are different from zero only
when the single particle indexes $l$ and $k$ belong to the subspaces
spanned by bases $\mathcal{B}_{0}$ and $\mathcal{B}_{1}$ and therefore
the complementary subspaces (of infinite dimension) are not required.
Finally, taking into account the unitarity \citep{ring2000} of the
matrices
\[
\bar{W}_{i}=\left(\begin{array}{cc}
\bar{U}_{i} & \bar{V}_{i}^{*}\\
\bar{V}_{i} & \bar{U}_{i}^{*}
\end{array}\right)
\]
it is possible to derive from Eq (\ref{eq:AB}) a set of identities
like $\bar{V}_{0}B+\bar{U}_{0}^{*}A=\left(\mathcal{R}_{1}^{T}\right)^{-1}\bar{U}_{1}^{*}$
that are essential to we arrive to the final result for the contractions
\begin{align}
\rho_{lk}^{01} & =\left[\bar{V}_{1}^{*}A^{-1}\bar{V}_{0}^{T}\right]_{lk}\label{eq:rho10F}\\
\bar{\kappa}_{k_{1}k_{2}}^{01} & =-\left[\left(\mathcal{R}^{T}\right)^{-1}\bar{U}_{1}^{*}A^{-1}\bar{V}_{0}^{T}\right]_{k_{1}k_{2}}\label{eq:kappa01F}\\
\kappa_{l_{1}l_{2}}^{10} & =\left[\bar{V}_{1}^{*}A^{-1}\bar{U}_{0}^{T}\left(\mathcal{R}^{T}\right)^{-1}\right]_{l_{1}l_{2}}\label{eq:kappa10F}
\end{align}
if the indexes belong to the subspaces spanned by bases $\mathcal{B}_{0}$
and $\mathcal{B}_{1}$ and zero otherwise. The matrix $A$, playing
a central role in the above expressions can be obtained from Eq (\ref{eq:AB})
and is given by
\begin{equation}
A=\bar{U}_{0}^{T}\left(\mathcal{R}^{T}\right)^{-1}\bar{U}_{1}^{*}+\bar{V}_{0}^{T}\mathcal{R}\bar{V}_{1}^{*}\label{eq:AFinal}
\end{equation}
For instance, the overlap of an one-body operator $\hat{O}=\sum_{ij}O_{kl}^{01}c_{0,k}^{\dagger}c_{1,l}$
with $O_{kl}^{01}=\mbox{}_{0}\langle k|\hat{O}|l\rangle_{1}$ is given
by $\mathrm{Tr}(O^{01}\rho^{01})$ in agreement with Eq (82) of I.
Please note that with the present formalism the formal developments
of Sec V of I leading from Eq (I.75) to Eq (I.82) are not required.
The new formulation presented in this paper does not affect the expression
for the overlap that is still given by Eq (I.58) 
\begin{equation}
\langle\phi_{0}|\phi_{1}\rangle=\sqrt{\det A\det\mathcal{R}}\label{eq:Overlap}
\end{equation}
This expression suffers from the sign indetermination of the square
root already present in the Onishi formula \citep{Onishi1966367}.
This indetermination can be resolved by using the pfaffian formula
for the overlap derived in Ref \citep{PhysRevC.79.021302}. The formula
obtained there was further generalized in Ref \citep{PhysRevC.84.014307}
to deal with the situation discussed here \textendash see Eqs (59-61)
of that reference. Later on, another, less general, pfaffian formula
for the overlap was given in Ref \citep{Avez2012}.

In the present derivation we have assumed that both bases $\mathcal{B}_{0}$
and $\mathcal{B}_{1}$ have the same dimensionality and the overlap
matrix $\mathcal{R}$ is a square invertible one. If this is not the
case and, for instance base $\mathcal{B}_{0}$ has a dimension $N_{0}$
smaller than $N_{1}$ (the dimension of $\mathcal{B}_{1}$) we can
complete $\mathcal{B}_{0}$ with $N_{1}-N_{0}$ orthogonal vectors
and assign occupancy 0 to them in the spirit of Eq (\ref{eq:block})
in order to get a square overlap matrix. 

The formulas can be further simplified by introducing the LU decomposition
of the overlap matrix $\mathcal{R}$
\[
\mathcal{R}=L_{0}^{*}L_{1}^{T}
\]
where $L_{0}$ and $L_{1}$ are lower triangular matrices. It introduces
a bi-orthogonal basis $|k)_{1}=\sum\left(L_{1}^{T}\right)_{jk}^{-1}|j\rangle_{1}$
and $_{0}(l|=\sum_{0}\langle i|\left(L_{0}^{*}\right)_{li}^{-1}$
such that $_{0}(l|k)_{1}=\delta_{lk}.$ The LU decomposition of the
overlap matrix suggests the definitions
\begin{align}
\tilde{U}_{0}=\left(L_{0}^{*}\right)^{-1}\bar{U}_{0}L_{0}^{+} & \qquad\tilde{V}_{0}=L_{0}^{+}\bar{V}_{0}L_{0}^{+}\label{eq:U0tilde}\\
\tilde{U}_{1}=\left(L_{1}^{*}\right)^{-1}\bar{U}_{1}L_{1}^{+} & \qquad\tilde{V}_{1}=L_{1}^{+}\bar{V}_{1}L_{1}^{+}\label{eq:U1tilde}
\end{align}
that allow to obtain quantities not depending explicitly on $\mathcal{R}$
like
\begin{equation}
\tilde{A}=\tilde{U}_{0}^{T}\tilde{U}_{1}^{*}+\tilde{V}_{0}^{T}\tilde{V}_{1}^{*}=L_{0}^{*}AL_{1}^{T}\label{eq:Atilde}
\end{equation}
The overlap is now written as
\begin{equation}
\langle\phi_{0}|\phi_{1}\rangle=\sqrt{\det\tilde{A}}\label{eq:Overtilde}
\end{equation}
It is also convenient to introduce the contractions
\begin{align}
\tilde{\rho}_{lk}^{01} & =\left[\tilde{V}_{1}^{*}\tilde{A}^{-1}\tilde{V}_{0}^{T}\right]_{lk}=L_{1}^{T}\rho^{01}L_{0}^{*}\label{eq:rho10F-1}\\
\tilde{\bar{\kappa}}_{k_{1}k_{2}}^{01} & =-\left[\tilde{U}_{1}^{*}\tilde{A}^{-1}\tilde{V}_{0}^{T}\right]_{k_{1}k_{2}}=L_{0}^{+}\bar{\kappa}{}^{01}L_{0}^{*}\label{eq:kappa01F-1}\\
\tilde{\kappa}_{l_{1}l_{2}}^{10} & =\left[\tilde{V}_{1}^{*}\tilde{A}^{-1}\tilde{U}_{0}^{T}\right]_{l_{1}l_{2}}=L_{1}^{T}\kappa{}^{01}L_{1}\label{eq:kappa10F-1}
\end{align}
Using them and the matrix elements $\tilde{O}=\left(L_{0}^{*}\right)^{-1}O^{01}\left(L_{\text{1}}^{T}\right)^{-1}$
\footnote{The matrix $\tilde{O}_{lk}$ is the one of the matrix elements of
the operator $\hat{O}$ in the bi-orthogonal basis $_{0}(l|$ and
$|k)_{1}$, i.e. $\tilde{O}_{lk}=_{0}(l|\hat{O}|k)_{1}=\sum\left(L_{0}^{*}\right)_{li}^{-1}{}_{0}\langle i|\hat{O}|j\rangle_{1}\left(L_{1}^{T}\right)_{jk}^{-1}$} one gets $\mathrm{Tr}(\tilde{O}\tilde{\rho}^{01})$ for the overlap
of an one-body operator. Similar considerations apply to the overlap
of two-body operators. Introducing the two-body matrix element in
the bi-orthogonal basis $\upsilon_{ijkl}^{B}=\mbox{}_{0}(ij|\hat{\upsilon}|kl)_{1}$
and related to $\upsilon_{ijkl}^{01}$ by 
\[
\upsilon^{B}=\left(L_{0}^{*}\right)^{-1}\left(L_{0}^{*}\right)^{-1}\upsilon^{01}\left(L_{1}^{T}\right)^{-1}\left(L_{1}^{T}\right)^{-1}
\]
we can define HF potential $\tilde{\Gamma}_{ik}^{01}=\frac{1}{2}\sum\tilde{\upsilon}_{ijkl}^{B}\tilde{\rho}_{lj}^{01}$
and pairing field $\tilde{\Delta}_{ij}^{01}=\frac{1}{2}\sum\tilde{\upsilon}_{ijkl}^{B}\tilde{\kappa}_{kl}^{01}$
to write 
\begin{equation}
\frac{\langle\phi_{0}|\hat{\upsilon}|\phi_{1}\rangle}{\langle\phi_{0}|\phi_{1}\rangle}=\frac{1}{2}\mathrm{Tr}[\tilde{\Gamma}^{01}\tilde{\rho}^{01}]-\frac{1}{2}\mathrm{Tr}[\tilde{\Delta}^{01}\tilde{\bar{\kappa}}^{01}]\label{eq:mv_v}
\end{equation}
which is again the standard expression but defined in terms of Eqs
(\ref{eq:rho10F-1}), (\ref{eq:kappa01F-1}) and (\ref{eq:kappa10F-1})
and the definitions above. The advantage of the definitions in Eqs
(\ref{eq:Atilde}), (\ref{eq:rho10F-1}), (\ref{eq:kappa01F-1}) and
(\ref{eq:kappa10F-1}) is that they have exactly the same expression
as the formulas available in the literature for complete basis but
expressed in terms of the ``tilde'' $U$ and $V$ matrices of Eqs
(\ref{eq:U0tilde}) and (\ref{eq:U1tilde}). There is an additional
advantage in the fact that $\tilde{A}$ is a ``more balanced'' matrix
being less affected by the near singular character of the overlap
matrix $\mathcal{R}.$ Let us finish by writing down the expression
of the density in coordinate space representation
\[
\rho^{01}(\vec{r})=\frac{\langle\phi_{0}|\hat{\rho}|\phi_{1}\rangle}{\langle\phi_{0}|\phi_{1}\rangle}=\sum_{ij}\varphi_{0i}^{*}(\vec{r})\varphi_{1j}(\vec{r})\rho_{ji}^{01}.
\]
often used along with zero range interactions.

Before finishing the presentation there are a few comments worth to
be mentioned
\begin{enumerate}
\item The simple form of the contractions of Eqs (\ref{eq:rho10F}), (\ref{eq:kappa01F})
and (\ref{eq:kappa10F}) and the fact that they are only different
from zero when the indexes belong to the subspaces spanned by bases
$\mathcal{B}_{0}$ and $\mathcal{B}_{1}$ is a direct consequence
of the definitions of Eqs (\ref{eq:rho01}), (\ref{eq:kappa01}) and
(\ref{eq:kappa10}) mixing single particle operators of both bases.
Those definitions are useful because we are expressing the operators
in the mixed form of Eq (\ref{eq:Twobody}).
\item The use of operators mixing creation and annihilation operators of
both bases as in Eq (\ref{eq:Twobody}) and the expressions of Eqs
(\ref{eq:rho01}), (\ref{eq:kappa01}) and (\ref{eq:kappa10}) were
already given in Ref \citep{BONCHE1990466} without proof and without
a justification of their interpretation as the contractions appearing
in the GWT.
\end{enumerate}

\section{Conclusions}

In this paper I have presented a modified version of the developments
of Ref \citep{PhysRevC.50.2874} that simplifies the application of
the generalized Wick's theorem for the calculation of operator overlaps
in the case of using two different nonequivalent bases for the two
HFB states entering the overlap. Applications of this formalism to
the case of harmonic oscillator bases with different oscillator lengths
will be discussed in a future publication.
\begin{acknowledgments}
This work has been supported by the Spanish Ministerio de Ciencia,
Innovación y Universidades and the European regional development fund
(FEDER), grants No PGC2018-094583-B-I00.
\end{acknowledgments}

\bibliographystyle{apsrev4-2}
\input Paper.bbl

\end{document}

%% file: Paper.bbl
%